\documentclass[aps,prb,twocolumn,superscriptaddress,showpacs,floatfix,longbibliography]{revtex4-2}

\usepackage[utf8]{inputenc}
\usepackage[T1]{fontenc}
\usepackage{amsmath,amssymb}
\usepackage{bm}
\usepackage{graphicx}
\usepackage{hyperref}
\usepackage{booktabs}
\usepackage{dcolumn}
\usepackage{xcolor}
\usepackage{enumitem}
\usepackage{placeins}

\hypersetup{
  colorlinks=true,
  linkcolor=blue!70!black,
  citecolor=green!50!black,
  urlcolor=blue!60!black,
}

\newcommand{\vR}{\mathbf{R}}
\newcommand{\vk}{\mathbf{k}}
\newcommand{\vd}{\mathbf{d}}
\newcommand{\vtau}{\boldsymbol{\tau}}

\newcommand{\fc}{f_{\mathrm{c}}}
\newcommand{\rcut}{r_{\mathrm{cut}}}
\newcommand{\rtaper}{r_{\mathrm{taper}}}
\newcommand{\RMSE}{\mathrm{RMSE}}
\newcommand{\nawf}{n_{\mathrm{awf}}}

\newcommand{\Nat}{N_{\mathrm{at}}}

\newcommand{\paoflow}{\textsc{paoflow}}

\begin{document}

\title{Environment-dependent tight-binding models from \textit{ab initio}
pseudo-atomic orbital Hamiltonians}

\author{M.~Buongiorno Nardelli}
\affiliation{Department of Physics, University of North Texas,
Denton, Texas 76203, USA}
\affiliation{Santa Fe Institute, Santa Fe, New Mexico 87501, USA}

\date{\today}

\begin{abstract}
\textit{Ab initio} pseudo-atomic orbital (PAO) Hamiltonians express the
electronic structure of a solid in a compact, localized basis that
spans the same Hilbert space as a conventional Slater--Koster
tight-binding model, thereby providing an exact \textit{ab initio}
representation without any loss of accuracy.
Building on this correspondence, we develop an environment-dependent
tight-binding (EDTB) framework in which Slater--Koster hopping integrals
are augmented with bond-screening functions that capture the local
coordination environment.  All parameters are determined by fitting to
the PAO eigenvalue spectrum across multiple atomic configurations
simultaneously, which breaks the degeneracy between screening and
hopping parameters and yields physically meaningful, transferable models
capable of generating Hamiltonians for large systems with \textit{ab initio}
precision.  We demonstrate the efficiency and accuracy of the approach on
four prototypical systems: bulk platinum, silicon surfaces, Si/Ge~[001]
superlattices, and twisted bilayer graphene with up to $4{,}324$ atoms.
The method is implemented in the \paoflow{} code and integrates seamlessly
with its full post-processing suite, enabling the evaluation of a broad range of electronic,
optical, and transport properties.
\end{abstract}

\maketitle

\section{Introduction}
\label{sec:intro}

The electronic structure of solids and nanostructures underpins a vast
range of physical phenomena---from charge transport and optical response
to topological order and correlated electron behavior.  First-principles
density-functional theory (DFT) provides a parameter-free description of
these properties, but its computational cost scales with system
size, limiting routine application to unit cells of at most a few hundred
atoms.  Tight-binding (TB) models offer a complementary approach: by
representing the Hamiltonian in a compact basis of localized orbitals,
they reduce the problem dimension by orders of magnitude while retaining
the essential orbital physics of the band
structure~\cite{SlaterKoster1954,Harrison1980,Papaconstantopoulos2015}. However, the method is intrinsically semi-empirical, since the TB parameters need to be fitted with existing experimental or theoretical data.

In recent years, we have developed a robust framework for constructing accurate, compact
Hamiltonians in a (pseudo) atomic orbital (PAO) basis (either from the pseudo-orbitals of the pseudopotential or from external atomic basis functions) directly from
plane-wave DFT calculations: the \paoflow{} code~\cite{BuongiornoNardelli2018,Cerasoli2020}
The main advantage of this representation is that the (pseudo) atomic orbital basis provides a direct matching with a TB-like model without any loss of accuracy, since the Hilbert space of both Hamiltonians share the same dimensionality. 
It is thus natural to integrate the \paoflow{} representation with the classic Slater-Koster TB framework and develop TB models with full \textit{ab initio} accuracy.

In the Slater--Koster (SK) framework~\cite{SlaterKoster1954}, hopping
integrals between atomic orbitals are decomposed into a small set of on-site energies and 
two-centers bond integrals $V_{ll'\mu}$ ($\mu = \sigma, \pi, \delta$) modulated by geometric
direction-cosine factors.  This two-center integrals are fitted to reproduce band structures from experiments or other theoretical methods and are bound to the equilibrium configuration used in the fitting. As such, 
their transferability to configurations with different local
environments---surfaces, interfaces, defects, or strained
geometries---is inherently limited.

Several strategies have been developed to extend the reach of TB models
beyond equilibrium bulk geometries. Among these are the rescaling methods originally proposed by Harrison~\cite{Harrison1980} and then extended by Goodwin, Skinner and Pettiford~\cite{Goodwin1989}, the Naval Research Laboratory (NRL)
tight-binding method~\cite{Cohen1994,Mehl1996,Bernstein2000,Papaconstantopoulos2015}
where the SK parameters are expressed as smooth functions of distance fitted to
DFT data across multiple crystal structures and volumes, thereby encoding
environment dependence implicitly through the training set, and the environment-dependent tight-binding (EDTB) formalism introduced by   
Tang, Wang, Chan, and Ho~\cite{Tang1996}, where each
hopping integral is multiplied by a screening factor that depends on a
bond screening sum---a measure of how crowded the local bonding
environment is around a given bond.

In this work we present a method
for constructing EDTB models capable of flexible screening granularity and distance-dependent hopping functions by fitting to the eigenvalue spectrum of
PAO Hamiltonians from \paoflow{}. Fitting directly to
PAO eigenvalues avoids the gauge ambiguity of Wannier-function
approaches~\cite{Urban2011} and leverages the orbital properties of the
PAO basis. The fitting is done using an efficient non-linear least-square Levenberg--Marquardt optimization with exact derivatives of the eigenvalue residuals via analytic Hellmann--Feynman Jacobians. Combined with optimized software design such as fully vectorized and parallelized algorithms and a block-sparse design-tensor formulation, we have developped a comprehensive EDTB code now integrated in the \paoflow{} sortware ecosystem. 

The paper is organized as follows.
Section~\ref{sec:paoflow} summarizes the \paoflow{} framework and the
PAO Hamiltonian construction.  Section~\ref{sec:sk} presents the
Slater--Koster formalism with environment-dependent screening and
distance-dependent hopping functions.  Section~\ref{sec:fitting}
describes the fitting procedure, including the objective function,
analytic Jacobian, regularization, and multi-geometry training protocol.
Section~\ref{sec:applications} presents applications to FCC platinum,
silicon surfaces, Si/Ge superlattices, and twisted bilayer graphene.
We summarize our conclusions in Section~\ref{sec:conclusions}.

\section{The PAOFLOW framework}
\label{sec:paoflow}

The reference electronic structure used throughout this work is provided
by the \paoflow{} code~\cite{BuongiornoNardelli2018,Cerasoli2020}, which
constructs a tight-binding-like Hamiltonian in a compact pseudo-atomic
orbital basis from a standard plane-wave DFT
calculation~\cite{Agapito2016, Agapito2016.1, Agapito2016.2}.

\subsection{Projection onto the PAO basis}
\label{sec:pao_projection}

Starting from the self-consistent Kohn--Sham eigenstates $|\psi_n\rangle$
obtained with a plane-wave code (e.g., Quantum
ESPRESSO~\cite{Giannozzi2009,Giannozzi2017}), one selects a set of $M$
orthonormal atomic-like orbitals $|\phi_\alpha\rangle$
($\langle\phi_\alpha|\phi_\beta\rangle = \delta_{\alpha\beta}$),
typically the L\"owdin-orthogonalized pseudo-atomic orbitals provided with
the pseudopotential.  These orbitals span a subspace $\mathcal{A}$ of
dimension $M$.  The projector onto $\mathcal{A}$ is
$\hat{P} = \sum_\alpha |\phi_\alpha\rangle\langle\phi_\alpha|$,
and the projection of each Bloch state yields
$|B_n\rangle = \hat{P}|\psi_n\rangle$ with coefficients
$B_{\alpha n} = \langle\phi_\alpha|\psi_n\rangle$.

The \emph{projectability} of each state,
\begin{equation}
  p_n = \langle\psi_n|\hat{P}|\psi_n\rangle
      = \sum_\alpha |B_{\alpha n}|^2,
\label{eq:proj}
\end{equation}
measures how completely $|\psi_n\rangle$ is described in $\mathcal{A}$.
States with $p_n$ below a threshold (typically $p_n <  0.95$) are
excluded from the Hamiltonian
reconstruction~\cite{Agapito2016}.

\subsection{Hamiltonian construction}
\label{sec:pao_ham}

From the $N$ retained states, one forms normalized projection vectors
$|A_n\rangle = |B_n\rangle/\sqrt{p_n}$ and constructs the PAO
Hamiltonian as~\cite{Agapito2016}
\begin{equation}
  \bar{H} = A\, E\, A^\dagger,
\label{eq:Hbar}
\end{equation}
where $E = \mathrm{diag}(\varepsilon_1,\ldots,\varepsilon_N)$ contains
the selected Kohn--Sham eigenvalues.  Since $\bar{H}$ is an
$M \times M$ matrix built from $N < M$ states, it possesses an
unphysical null space of dimension $M - N$.  This is removed by adding a
shift projector~\cite{Agapito2016}:
\begin{equation}
  \bar{H}_\kappa = \bar{H} + \kappa\, Q_{\mathcal{N}},
  \quad
  Q_{\mathcal{N}} = I_M - A(A^\dagger A)^{-1} A^\dagger,
\label{eq:Hkappa}
\end{equation}
which pushes the null-space eigenvalues to an energy $\kappa$ above the
physical bands, leaving the projected eigenstates unaffected.

The PAO Hamiltonian provides an exact description of the electronic properties of the
crystal without any iterative optimization of the basis functions, and maps directly onto a corresponding SK Hamiltonian with the same orbital basis (angular momentum labels)~\cite{BuongiornoNardelli2018,Cerasoli2020}.  The real-space
representation $H_{\alpha\beta}(\vR)$, obtained by inverse Fourier
transform on the DFT $k$-mesh, decays rapidly with $|\vR|$.  The Bloch
Hamiltonian at arbitrary wave vector $\vk$ is reconstructed by
\begin{equation}
  H_{\alpha\beta}(\vk)
  = \sum_{\vR} e^{i\vk\cdot\vR}\, H_{\alpha\beta}(\vR),
\label{eq:fourier}
\end{equation}
and diagonalization yields the band structure
$\varepsilon_{n\vk}^{\mathrm{PAO}}$.

\section{Slater--Koster tight-binding with environment dependence}
\label{sec:sk}

\subsection{Two-center hopping integrals}
\label{sec:sk_hopping}

In the SK scheme~\cite{SlaterKoster1954}, the hopping matrix element
between orbital $\alpha$ on atom $i$ and orbital $\beta$ on atom $j$,
connected by a bond vector
$\vd_{ij} = \vR + \vtau_j - \vtau_i$ with $|\vd_{ij}| = d_{ij}$, is
\begin{equation}
  H_{\alpha\beta}^{\mathrm{SK}}(\vd_{ij})
  = \sum_\mu c_{\alpha\beta\mu}(\hat{\vd}_{ij})\,
    V_{l_\alpha l_\beta \mu}^{(s)},
\label{eq:sk_hop}
\end{equation}
where $c_{\alpha\beta\mu}(\hat{\vd})$ are direction-cosine factors,
$l_\alpha$, $l_\beta$ are angular-momentum quantum numbers,
$\mu \in \{\sigma, \pi, \delta\}$ labels the bond symmetry, and
the superscript $(s)$ indicates the neighbor shell.
For an $spd$ basis, there are ten independent bond integrals per shell:
$V_{ss\sigma}$, $V_{sp\sigma}$, $V_{pp\sigma}$, $V_{pp\pi}$,
$V_{sd\sigma}$, $V_{pd\sigma}$, $V_{pd\pi}$, $V_{dd\sigma}$,
$V_{dd\pi}$, $V_{dd\delta}$.

\subsection{Design-tensor representation}
\label{sec:design_tensor}

The linearity of $H_{\alpha\beta}^{\mathrm{SK}}$ in the bond integrals
motivates the introduction of a design tensor
$M_{b\lambda\alpha\beta}^{(s)}$, where $b$ indexes bonds in shell $s$
and $\lambda$ indexes the active SK parameters:
\begin{equation}
  M_{b\lambda\alpha\beta}^{(s)}
  = c_{\alpha\beta\mu(\lambda)}\!\left(\hat{\vd}_b\right).
\end{equation}
The full SK Bloch Hamiltonian can thus be written as
\begin{equation}
  H_{\alpha\beta}^{\mathrm{SK}}(\vk)
  = \sum_p \varepsilon_p\, \Omega_{p,\alpha\beta}
  + \sum_{s=1}^{N_s} \sum_b e^{i\vk\cdot\vR_b}
    \sum_\lambda V_\lambda^{(s)}\, M_{b\lambda\alpha\beta}^{(s)},
\label{eq:sk_bloch}
\end{equation}
where $\Omega_{p,\alpha\beta}$ is a sparse on-site map tensor.
The design tensor inherits a pronounced block-sparse structure: a bond
connecting atoms $i$ and $j$ produces nonzero entries only in the
orbital sub-block $\alpha \in \mathcal{O}_i$,
$\beta \in \mathcal{O}_j$.  We exploit this by grouping bonds into
bond groups $\mathcal{G}_{ij}$ that share the same atom pair and
operating on compact sub-blocks, yielding significant speedups for
multi-atom unit cells.

\subsection{Environment-dependent screening}
\label{sec:screening}

In the EDTB extension~\cite{Tang1996}, each two-center hopping integral
is modulated by an environment-dependent screening factor:
\begin{equation}
  \tilde{V}_\lambda^{(s)}(i,j)
  = V_\lambda^{(s)}\,
    \exp\!\left(-\gamma_\lambda\, S_{ij}\right),
\label{eq:screened_hop}
\end{equation}
where $S_{ij}$ is a bond screening sum and $\gamma_\lambda \geq 0$ is
a screening strength parameter.  The screening sum
\begin{equation}
  S_{ij} = \sum_{k \neq i,j} \fc(d_{ik})\, \fc(d_{jk})
\label{eq:screen_sum}
\end{equation}
runs over all atoms $k$ that are simultaneously close to both bond
endpoints, with $\fc$ a smooth cosine-tapered cutoff function:
\begin{equation}
  \fc(d) =
  \begin{cases}
    1 & d \leq \rtaper, \\[3pt]
    \tfrac{1}{2}\bigl[1 + \cos\!\bigl(
      \pi\,\tfrac{d - \rtaper}{\rcut - \rtaper}\bigr)\bigr]
    & \rtaper < d < \rcut, \\[4pt]
    0 & d \geq \rcut,
  \end{cases}
\label{eq:cutoff}
\end{equation}
where $\rcut$ is the cutoff radius and $\rtaper = 0.8\,\rcut$.
The physical content of Eq.~(\ref{eq:screened_hop}) is transparent:
the more atoms surround a bond, the larger $S_{ij}$ becomes, and the
more strongly the hopping is screened.  This mechanism captures, at the
two-center level, the modification of covalent bonding by the local
coordination environment.

The screening strengths $\gamma_\lambda$ can be parameterized at
three levels of granularity:
(i)~a single global $\gamma$ shared by all channels (1~parameter);
(ii)~one $\gamma_{ll'}$ per angular-momentum pair (up to
6~parameters for $spd$);
(iii)~one $\gamma_\mu$ per independent SK bond integral
(up to 10~parameters).
The EDTB Bloch Hamiltonian thus reads
\begin{equation}
\begin{split}
  H_{\alpha\beta}^{\mathrm{EDTB}}(\vk)
  &= \sum_p \varepsilon_p\, \Omega_{p,\alpha\beta}
   + \sum_{s,b} e^{i\vk\cdot\vR_b} \\
  &\quad \times  \sum_\lambda V_\lambda^{(s)} 
    e^{-\gamma_{q(\lambda)} S_b}\,
    M_{b\lambda\alpha\beta}^{(s)},
\end{split}
\label{eq:edtb_bloch}
\end{equation}
where $q(\lambda)$ maps each hopping parameter to the appropriate
screening index.

Optionally, on-site energies can be made environment-dependent through
coordination-dependent shifts:
\begin{equation}
  \tilde{\varepsilon}_\alpha
  = \varepsilon_\alpha + \eta_\alpha\, C_i,
  \quad
  C_i = \sum_{k \neq i} \fc(d_{ik}),
\label{eq:onsite_shift}
\end{equation}
where $\eta_\alpha$ is a shift parameter and $C_i$ is the effective
coordination number of atom~$i$.

\subsection{Distance-dependent hopping functions}
\label{sec:dd_model}

The discrete-shell model assigns independent hopping parameters
$V_\lambda^{(s)}$ to each shell.  For systems with continuously
varying bond lengths we replace the shell-resolved parameters with a
Goodwin--Skinner--Pettifor functional form~\cite{Goodwin1989}:
\begin{equation}
\begin{split}
  V_\lambda(r)
  &= V_{0,\lambda}\,
    \biggl(\frac{r_0}{r}\biggr)^{\!n_\lambda} \\
  &\quad\times
    \exp\!\Biggl[
      n_\lambda\!\biggl(
        \!-\!\Bigl(\frac{r}{r_c}\Bigr)^{\!n_c}
        \!+\!\Bigl(\frac{r_0}{r_c}\Bigr)^{\!n_c}
      \biggr)
    \Biggr],
\end{split}
\label{eq:goodwin}
\end{equation}
where $V_{0,\lambda}$ is the hopping amplitude at the reference distance
$r_0$, $n_\lambda > 0$ is a per-channel decay exponent, $r_c$ is the
cutoff radius, and $n_c > 0$ is a shared cutoff steepness.  As $r$
increases, the exponential factor ensures a rapid and smooth decay to
zero, and $V_\lambda(r_0) = V_{0,\lambda}$ by construction.

The screened distance-dependent Bloch Hamiltonian becomes
\begin{equation}
\begin{split}
  H_{\alpha\beta}^{\mathrm{DD}}(\vk)
  &= \sum_p \varepsilon_p\, \Omega_{p,\alpha\beta}
   + \sum_b e^{i\vk\cdot\vR_b}\\
  &\quad\times \sum_\lambda V_\lambda(d_b)\,
    e^{-\gamma_{q(\lambda)} S_b}\,
    M_{b\lambda\alpha\beta}.
\end{split}
\label{eq:dd_bloch}
\end{equation}

A natural initialization for the distance-dependent fit is obtained from
an existing discrete-shell fit: for each channel $\lambda$,
$V_\lambda^{(1)}$ sets $V_{0,\lambda}$, and the decay exponent
$n_\lambda$ is estimated from the ratio of first- and second-shell
hoppings.

\subsection{Multi-parameter model for complex geometries}
\label{sec:multiparam}

To model systems with chemically distinct environments---surfaces,
interfaces, or multi-species systems such as alloys or heterostructures---we generalize the EDTB
Hamiltonian to a multi-parameter framework.  Each atom is assigned an
environment label $\ell_i$ (e.g., ``bulk,'' ``surface,'' or a species
identifier), and the hopping parameters for a bond connecting atoms
$i$ and $j$ are selected as
\begin{equation}
  V_{ij}^{(\mathrm{bond})} =
  \begin{cases}
    V^{(\ell_i)} & \text{if } \ell_i = \ell_j, \\[3pt]
    \mathcal{M}(V^{(\ell_i)}, V^{(\ell_j)})
      & \text{if } \ell_i \neq \ell_j,
  \end{cases}
\label{eq:multiparam}
\end{equation}
where $\mathcal{M}$ denotes a mixing operator for interface bonds.
We employ by default the geometric mean,
$\mathcal{M}_{\mathrm{geo}}(a,b) = \mathrm{sgn}(a+b)\,\sqrt{|ab|}$,
which preserves the sign structure of the SK integrals.  Arithmetic
and first-label mixing are also available.  On-site energies are
selected by each atom's own label and chemical species.

This multi-parameter framework enables the construction of
Hamiltonians for systems substantially larger than the DFT training
cells---for instance, thick slabs, long-period superlattices, or
multi-component heterostructures---by combining a small number of
independently or jointly fitted parameter sets, each trained on a
manageable unit cell, and assembling them according to the
environment labels and mixing rules defined above.  Because the
screening functions and hopping parameters are transferable across
local environments, no additional DFT calculations are required for
the target system.

\section{Fitting procedure}
\label{sec:fitting}

\subsection{Objective function}
\label{sec:objective}

The model parameters $\mathbf{p}$ are determined by minimizing the
root-mean-square eigenvalue error over a uniform Brillouin-zone mesh:
\begin{equation}
  \RMSE(\mathbf{p})
  = \biggl[\frac{1}{N_k N_b}
    \sum_{\vk} \sum_{n=1}^{N_b}
    \bigl(\varepsilon_{n\vk}^{\mathrm{SK}}
        - \varepsilon_{n\vk}^{\mathrm{PAO}}\bigr)^{\!2}
    \biggr]^{\!1/2}\!.
\label{eq:rmse}
\end{equation}
Similarly to other TB fitting implementations~\cite{Kim2018TBFIT,Nakhaee2020,Hegde2014}, minimization is performed with the Levenberg--Marquardt
algorithm~\cite{Levenberg1944,Marquardt1963}, which requires the
Jacobian matrix $J_{m,j} = \partial r_m / \partial p_j$ of the
residual vector.

\subsection{Analytic Jacobian via the Hellmann--Feynman theorem}
\label{sec:jacobian}

The derivative of each eigenvalue with respect to any parameter $p_j$
is computed analytically from the Hellmann--Feynman
theorem:
\begin{equation}
  \frac{\partial \varepsilon_{n\vk}}{\partial p_j}
  = \langle n\vk | \frac{\partial H(\vk)}{\partial p_j} | n\vk \rangle.
\label{eq:hf}
\end{equation}
For on-site energies,
$\partial \varepsilon_{n\vk}/\partial \varepsilon_p
= \sum_{\alpha \in p} |\psi_{n\vk}^\alpha|^2$.
For hopping parameters,
\begin{equation}
  \frac{\partial H_{\alpha\beta}}{\partial V_\lambda^{(s)}}
  = \sum_b e^{i\vk\cdot\vR_b}\,
    e^{-\gamma_{q(\lambda)} S_b}\,
    M_{b\lambda\alpha\beta}^{(s)}.
\end{equation}
For screening strengths,
\begin{equation}
\begin{split}
  \frac{\partial H_{\alpha\beta}}{\partial \gamma_q}
  &= -\sum_{s,b} e^{i\vk\cdot\vR_b}
    \sum_{\lambda: q(\lambda)=q} \\
  &\quad\times
    V_\lambda^{(s)}\, S_b\,
    e^{-\gamma_q S_b}\,
    M_{b\lambda\alpha\beta}^{(s)}.
\end{split}
\end{equation}
Analogous expressions hold for the distance-dependent parameters
$V_{0,\lambda}$, $n_\lambda$, and $n_c$, obtained via the chain rule
applied to Eq.~(\ref{eq:goodwin}).

The analytic Jacobian avoids finite-difference approximations entirely,
providing both speed and numerical stability.  All derivatives are
evaluated from the same eigenvector matrix used for the forward model,
at negligible additional cost.

Alternative optimization strategies have been employed for TB parameter
fitting, including genetic
algorithms~\cite{Hegde2010,Phan2025,Froseth2003} and
active-learning-driven iterative
schemes~\cite{GarrityThreeBody2023}.  Derivative-free methods such as
genetic algorithms or simulated annealing are attractive for their
ability to explore rugged landscapes and avoid local minima, but they
typically require $10^{4}$--$10^{6}$ objective-function evaluations and
scale poorly with the number of parameters.  Active-learning
approaches~\cite{GarrityThreeBody2023} iteratively expand the training
set by testing model predictions against new DFT calculations,
effectively addressing the data-generation problem rather than the
optimization itself.  Our approach is complementary: the availability of
analytic Jacobians makes each Levenberg--Marquardt iteration
inexpensive, so convergence is reached in $\mathcal{O}(10^{2})$
iterations even for 30--40 parameters.  We compensate for the
susceptibility of gradient-based methods to local minima through
multi-start optimization, which remains efficient because each
individual run is fast.

\subsection{Regularization}
\label{sec:regularization}

To suppress unphysical oscillations in the hopping parameters---a strategy
also adopted in recent TB parameterization
work~\cite{Wang2021ML_TB,Kim2023,Phan2025}---we
optionally augment the residual vector with Tikhonov penalty terms:
\begin{equation}
  \tilde{r}_m(\mathbf{p}) =
  \begin{cases}
    r_m(\mathbf{p}) & m = 1,\ldots,N_k N_b, \\
    \alpha\, w_j\, p_j & m = N_k N_b + j,
  \end{cases}
\end{equation}
where $\alpha$ is the regularization strength and $w_j$ are
parameter-class weights.

\subsection{Multi-geometry training}
\label{sec:multigeom}

For a single crystal geometry, the screening parameters
$\gamma_\lambda$ are partially redundant with the shell-dependent
hopping integrals: the uniform coordination in a perfect crystal
renders $S_{ij}$ identical for all symmetry-equivalent bonds, so any
constant screening can be absorbed into the hoppings.  The power of the
EDTB formalism emerges when a single shared parameter set is fitted
simultaneously to band structures from $N_g$ configurations with
different coordination environments~\cite{Papaconstantopoulos2015}.

The combined residual vector is
\begin{equation}
  \tilde{\mathbf{r}}(\mathbf{p})
  = \begin{pmatrix}
      w_1\, \mathbf{r}^{(1)}(\mathbf{p}) \\
      \vdots \\
      w_{N_g}\, \mathbf{r}^{(N_g)}(\mathbf{p})
    \end{pmatrix},
\label{eq:multigeom}
\end{equation}
where $w_g$ are per-geometry weights.  The Jacobian is assembled by
vertical concatenation of the per-geometry Jacobians.

We employ the following fitting protocol: (1)~a standard SK fit
(without screening) at the equilibrium geometry to obtain initial
two-center parameters; (2)~a multi-geometry EDTB fit initialized with
the stage-1 parameters and small random screening values.  Multi-start
optimization (multiple independent Levenberg--Marquardt runs from
random initial points) is used to guard against local minima.

\section{Applications}
\label{sec:applications}

\subsection{FCC platinum}
\label{sec:pt}

As a first demonstration, we consider FCC platinum with an $spd$ basis
(9~orbitals per atom) and three neighbor shells and an equilibrium lattice
parameter $a_0 = 3.93$~\AA.  The PAO Hamiltonian is constructed with \paoflow{} from a
$12^3$ $k$-mesh.  Self-consistent DFT calculations are performed at seven
isotropic volumes spanning strains
$\varepsilon \in \{-4\%, -3\%, -2\%, 0\%, +1\%, +2\%, +4\%\}$.  Five
volumes ($\varepsilon = -4\%, -2\%, 0\%, +2\%, +4\%$) constitute
the training set; two ($\varepsilon = -3\%, +1\%$) are held out for
blind testing.

We compare three models, all sharing the same orbital basis:
(i)~SK-eq: a standard SK fit trained only at equilibrium (34~parameters:
4~on-site energies, $30$~hoppings);
(ii)~EDTB-eq: an EDTB fit with per-$ll'$ screening, also trained at
equilibrium only (40~parameters);
(iii)~EDTB-multi: an EDTB fit trained simultaneously on all five training
volumes (40~parameters).

\begin{figure*}[tp]
  \centering
  \includegraphics[width=\textwidth]{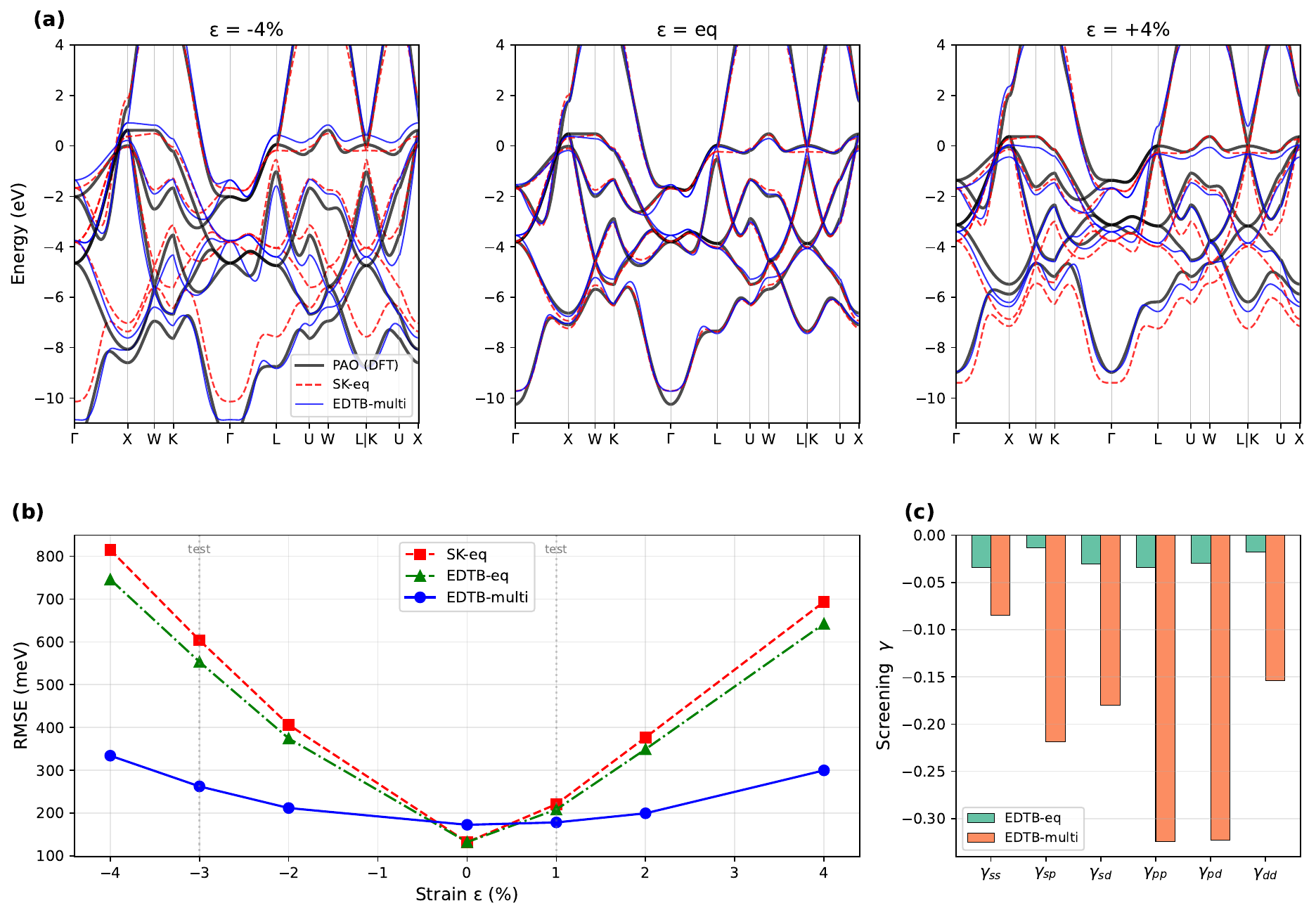}
  \caption{FCC platinum volume-scan results.
  (a)~Band structures at $\varepsilon = -4\%$ (left), equilibrium
  (center), and $+4\%$ (right); black: PAO reference, red dashed:
  SK-eq, blue: EDTB-multi.
  (b)~RMSE vs.\ strain for the three models; grey lines mark the
  test volumes.
  (c)~Fitted screening strengths $\gamma_{ll'}$ for EDTB-eq (green)
  and EDTB-multi (orange).}
  \label{fig:pt_results}
\end{figure*}

\begin{table}[t]
\centering
\caption{Per-volume RMSE (meV) for FCC platinum.  ``Train'' and ``test''
indicate whether the volume was used in the multi-geometry fit.}
\label{tab:rmse}
\begin{tabular}{llD{.}{.}{0}D{.}{.}{0}D{.}{.}{0}}
  \toprule
  Strain & Type & \multicolumn{1}{c}{SK-eq} &
    \multicolumn{1}{c}{EDTB-eq} & \multicolumn{1}{c}{EDTB-multi} \\
  \midrule
  $-4\%$ & train & 815 & 746 & 334 \\
  $-3\%$ & test  & 604 & 553 & 262 \\
  $-2\%$ & train & 406 & 375 & 212 \\
    eq   & train & 132 & 132 & 173 \\
  $+1\%$ & test  & 221 & 208 & 178 \\
  $+2\%$ & train & 377 & 349 & 200 \\
  $+4\%$ & train & 694 & 642 & 300 \\
  \midrule
  \multicolumn{2}{l}{Combined (RMS)} & 519 & 478 & 244 \\
  \multicolumn{2}{l}{Test only}      & 455 & 418 & 224 \\
  \midrule
  \multicolumn{2}{l}{\# Parameters}  & 34  & 40  & 40  \\
  \bottomrule
\end{tabular}
\end{table}

Figure~\ref{fig:pt_results}(a) compares the band structures and shows how at $\varepsilon = \pm 4\%$, visible
deviations appear in the SK-eq $d$-band manifold, while the EDTB-multi model tracks the PAO reference closely
across the entire Brillouin zone.  Figure~\ref{fig:pt_results}(b) shows the RMSE vs. strain data from Table~\ref{tab:rmse} while Figure~\ref{fig:pt_results}(c) displays the screening $\gamma$'s for the two EDTB fitting procedures. The fitted screening strengths from
the multi-geometry fit display a clear angular-momentum hierarchy
($|\gamma_{pp}| \approx |\gamma_{pd}| > |\gamma_{sp}| > |\gamma_{sd}|
> |\gamma_{dd}| > |\gamma_{ss}|$), with the largest values for
$p$--$p$ and $p$--$d$ interactions, which are the most sensitive to
coordination changes.  These physically meaningful screening
strengths emerge only when the optimizer is forced to fit multiple
volumes simultaneously.

\begin{figure*}[tp]
  \centering
  \includegraphics[width=\textwidth]{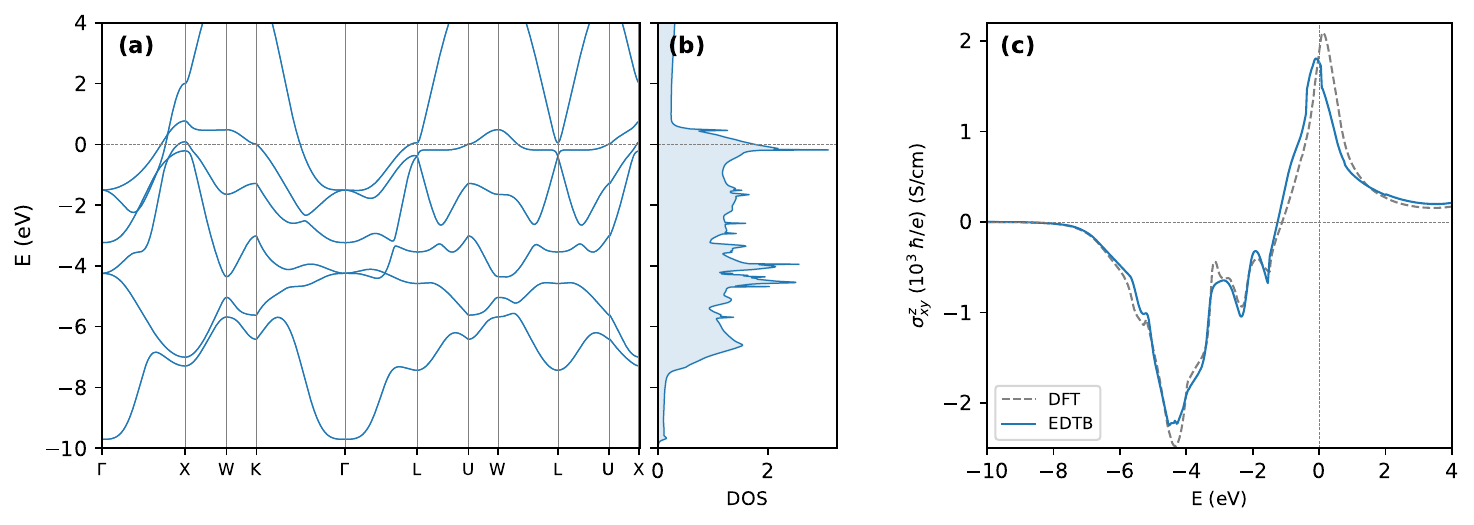}
  \caption{EDTB model for FCC Pt with ad-hoc SOC, processed through the
    full \paoflow{} pipeline.  (a)~Band structure showing SOC-induced
    splittings.  (b)~Density of states.
    (c)~Intrinsic spin Hall conductivity $\sigma_{xy}^z(E)$.}
  \label{fig:pt_soc_shc}
\end{figure*}

Table~\ref{tab:rmse} reports the RMSE at each volume.  At equilibrium,
all three models achieve comparable accuracy ($\sim$130--170~meV).
Away from equilibrium, the models diverge sharply: the SK-eq RMSE
reaches 815~meV at $-4\%$ strain, while the EDTB-multi model
maintains sub-350~meV accuracy across the entire $\pm 4\%$ range.  The
combined RMSE is reduced from 519~meV (SK-eq) to 244~meV---a factor of
two improvement.  Crucially, the two test volumes show the same trend
(test RMSE reduced from 455~meV to 224~meV), confirming genuine
transferability rather than overfitting.

A key advantage of the EDTB parameterization is that the resulting
Hamiltonian is fully compatible with the \paoflow{} post-processing
pipeline.  To illustrate this, we augment the EDTB-multi model with
\emph{ad hoc} atomic spin--orbit coupling (SOC),
\begin{equation}
  H_{\mathrm{SOC}} = \sum_{i,l} \xi_l^{(i)}\,
    \mathbf{L}_i \cdot \mathbf{S}_i,
\label{eq:soc}
\end{equation}
using $\xi_p = 0.553$~eV for the dominant $p$-channel contribution in
Pt~\cite{Buongiorno2024}.  The SOC doubles the Hilbert-space dimension
and lifts band degeneracies throughout the Brillouin zone, as seen in Figure \ref{fig:pt_soc_shc}(a,b) for the band structure and density of states.

Having a fully SOC Hamiltonian allows us to calculate, for instance, the intrinsic spin Hall conductivity (SHC) from the Kubo formula~\cite{Guo2005,BuongiornoNardelli2018,Cerasoli2020},
\begin{equation}
  \sigma_{xy}^z = \frac{e}{\hbar} \sum_{n} \int_{\mathrm{BZ}}
  \frac{d^3k}{(2\pi)^3}\, f_{n\vk}\,
  \Omega_{n,xy}^{z}(\vk),
\label{eq:kubo_shc}
\end{equation}
where $\Omega_{n,xy}^{z}(\vk)$ is the spin Berry curvature.
Figure~\ref{fig:pt_soc_shc}(c) shows the resulting SHC as a function of Fermi energy.
These results are in good agreement with first-principles relativistic
calculations~\cite{Guo2008,Qiao2018,BuongiornoNardelli2018,Cerasoli2020}, demonstrating that the EDTB model
augmented with ad-hoc SOC captures the essential physics of
spin--orbit-driven transport.

\subsection{Silicon surfaces}
\label{sec:si_surfaces}

The silicon EDTB model uses an $spd$ basis (9~orbitals/atom) with
three neighbor shells.  The model is trained simultaneously on four
geometries: bulk diamond Si ($6^3$ $k$-mesh, weight $w=2.0$) and three
unrelaxed, unpassivated 8-atom slabs for the (001), (110), and (111)
surfaces ($6{\times}6{\times}1$ $k$-mesh, $w=1.0$ each).  The elevated
bulk weight ensures accurate bulk band structure reproduction, while the
surface slabs provide the coordination variation needed to constrain the
screening.  The total parameter count is 40 (34~SK + 6~screening).

The multi-geometry training yields screening strengths of
significantly larger magnitude compared to bulk-only fitting---in
particular, the $sd$ and $pd$ screening strengths increase by more than
an order of magnitude, indicating that $d$-orbital hybridization at the
surface is substantially modified by the reduced coordination.

To build thick slabs larger than the DFT training cells, we adopt a
dual-parameter approach: each atom is classified as ``bulk'' or
``surface'' based on its coordination, and the Hamiltonian matrix
elements are constructed using the fitted EDTB parameters with
the local $S_{ij}$ values.  For bonds bridging
different environments, hopping parameters are mixed using the
geometric-mean rule.  We apply this model to 40-atom slabs for each
surface orientation.

\begin{figure*}[tp]
  \centering
  \includegraphics[width=\textwidth]{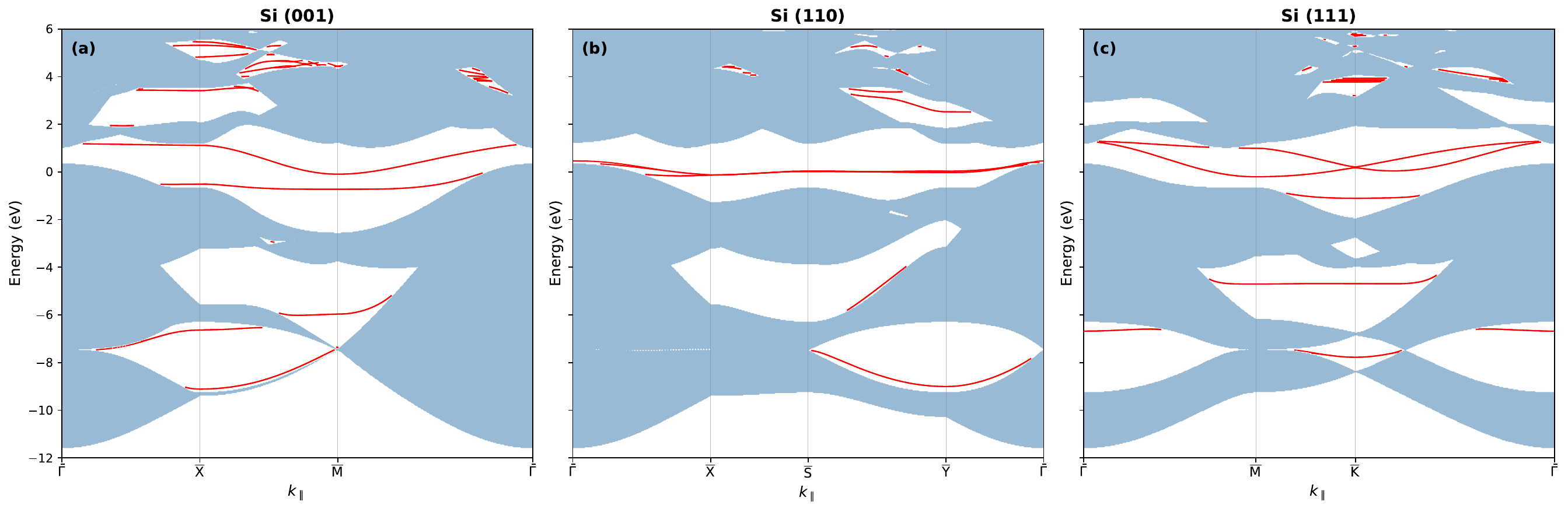}
  \caption{Projected bulk band structure (blue shaded) and surface
  states (red) for three silicon surfaces: (a)~Si(001),
  (b)~Si(110), (c)~Si(111) 40-atom unrelaxed, unpassivated slabs.}
  \label{fig:si_surfaces}
\end{figure*}

Figure~\ref{fig:si_surfaces} shows the results for all three low-index
surfaces.  The projected bulk bands (shaded regions) correctly reproduce
the expected bulk gaps, and slab states in the projected gaps are observed
for all orientations.  While the unrelaxed, unpassivated slab geometries
preclude quantitative comparison with experiment, this calculation
demonstrates that the EDTB multi-geometry framework combined with
dual-parameter slab construction provides a fast, physically transparent
approach to modeling surface electronic structure.

\subsection{Si/Ge~[001] superlattices}
\label{sec:sige}

As a test of the multi-parameter framework for binary 
heterostructures, we consider Si/Ge~[001] superlattices.
An EDTB model is first fitted to a 16-atom
Si$_8$Ge$_8$ training cell using an $spd$ basis (9~orbitals/atom) with
3~neighbor shells and per-$ll'$ screening.  Each atom in the prediction
superlattice is classified as Si$_{\mathrm{bulk}}$,
Ge$_{\mathrm{bulk}}$, or interface (any atom with a cross-species nearest
neighbor), and the multi-parameter Hamiltonian is assembled using
geometric-mean mixing at the interface.

\begin{figure}[t]
  \centering
  \includegraphics[width=\columnwidth]{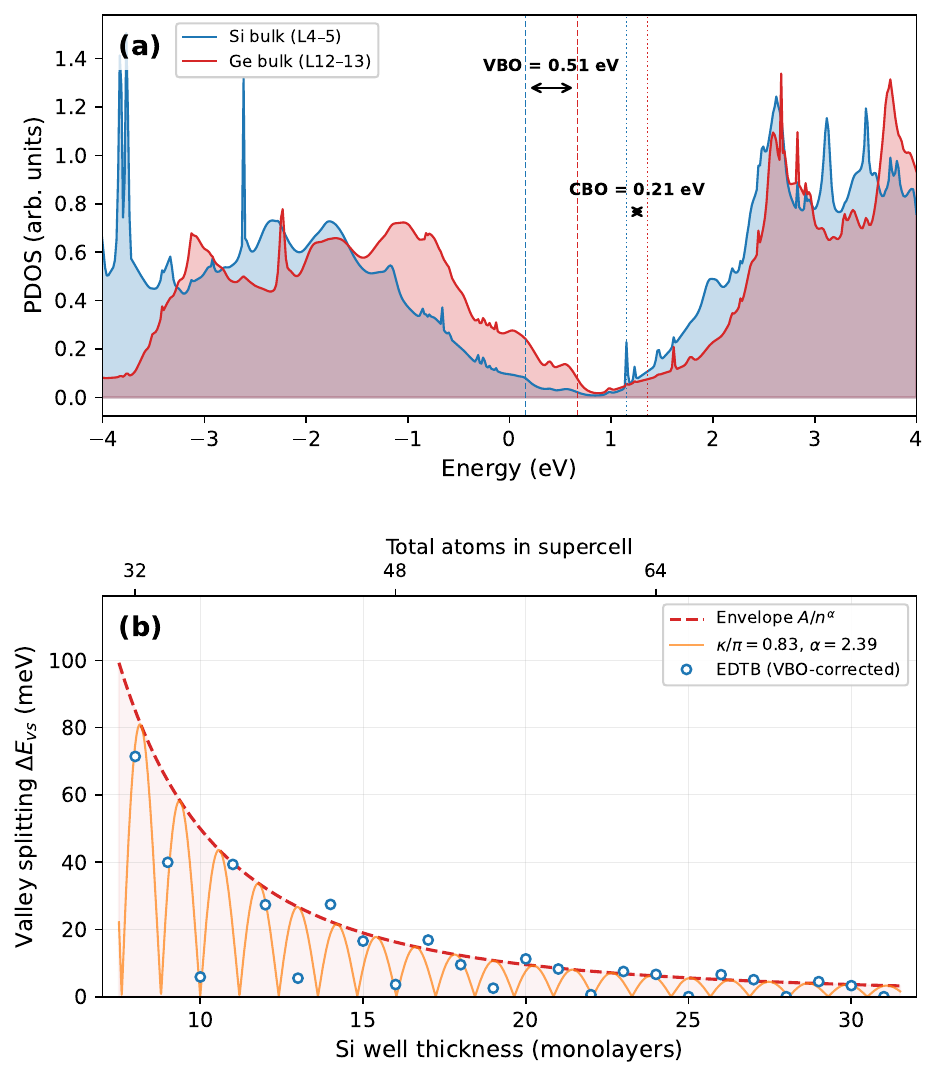}
  \caption{Band offsets and valley splitting in Si/Ge~[001]
  superlattices.
  (a)~Layer-resolved PDOS for bulk-like Si (blue) and Ge (red) layers
  of the Si$_{16}$Ge$_{16}$ superlattice, with extracted VBM (dashed)
  and CBM (dotted) positions indicating the valence band offset
  (VBO~$\approx 0.5$~eV) and conduction band offset
  (CBO~$\approx 0.2$~eV).
  (b)~Valley splitting $\Delta E_{\mathrm{vs}}$ vs Si well thickness
  for Si$_n$/Ge$_8$ ($n = 8$--31~ML).  Open circles: EDTB data;
  solid curve: fit to Eq.~(\ref{eq:valley_splitting}); dashed line:
  power-law envelope.  The oscillation period
  $\kappa/\pi \approx 0.83$ matches the bulk Si $\Delta$-valley
  position.}
  \label{fig:sige}
\end{figure}

Realistic band offsets are introduced through rigid on-site energy shifts
applied to the Ge orbitals:
$\Delta\varepsilon_p^{\mathrm{Ge}} = \mathrm{VBO}_{\mathrm{target}} =
1.05$~eV and
$\Delta\varepsilon_s^{\mathrm{Ge}} = \mathrm{CBO}_{\mathrm{target}} =
0.0$~eV, chosen to reproduce the valence band alignment of Ge strained
to the Si lattice constant.  Layer-resolved projected densities of states
for the 32-atom Si$_{16}$Ge$_{16}$ superlattice
[Fig.~\ref{fig:sige}(a)] display a valence band offset
$\mathrm{VBO} = \mathrm{VBM}_{\mathrm{Si}} -
\mathrm{VBM}_{\mathrm{Ge}} \approx 0.5$~eV
and a conduction band offset
$\mathrm{CBO} = \mathrm{CBM}_{\mathrm{Si}} -
\mathrm{CBM}_{\mathrm{Ge}} \approx 0.2$~eV,
consistent with the expected type-II band alignment at the strained
Si/Ge~[001] interface.  The spatial separation of Si-like and Ge-like
electronic states confirms that the multi-parameter framework correctly
propagates the on-site energy differences into the superlattice
Hamiltonian.

An important observable in Si/Ge heterostructures is the valley splitting
$\Delta E_{\mathrm{vs}}$---the energy difference between the two lowest
conduction-band states at $\Gamma$ that descend from the bulk Si
$\Delta$-valley pair folded onto the zone center by the [001]
periodicity.  We compute $\Delta E_{\mathrm{vs}}$ for a systematic
series of Si$_n$/Ge$_8$ superlattices with Si well thickness
$n = 8$--31~monolayers~(ML) in steps of 1~ML, covering systems up to
78~atoms.  The Ge barrier is fixed at 8~ML; when the total monolayer
count is not divisible by four, additional Ge monolayers are appended to
close the supercell, with negligible effect on the splitting.

The computed valley splitting [Fig.~\ref{fig:sige}(b)] oscillates
rapidly with well width and decays with a power-law envelope, consistent
with established results~\cite{BoykinPRB2004,FriesenPRB2007}.  A
nonlinear least-squares fit to
\begin{equation}
  \Delta E_{\mathrm{vs}}(n) =
    \frac{A}{n^{\alpha}}\,
    \bigl|\cos(\kappa\, n + \varphi)\bigr|
  \label{eq:valley_splitting}
\end{equation}
yields envelope exponent $\alpha = 2.4 \pm 0.4$ and oscillation wave
vector $\kappa/\pi = 0.83 \pm 0.01$ ($R^2 = 0.80$).  The oscillation
period matches the position of the bulk Si $\Delta$-valley minimum at
$\sim 0.85\,(2\pi/a)$ along $\Gamma$--X, confirming that the EDTB model
correctly captures the zone-folding origin of the splitting.  The
power-law exponent $\alpha \approx 2.4$ reflects the rapid suppression
of intervalley coupling with increasing well width, in qualitative
agreement with the expected trend for sharp-interface
heterostructures~\cite{BoykinPRB2004}.

This application demonstrates that the multi-parameter EDTB framework
extends naturally to binary heterostructures: the combination of a single
pre-fit on a small training geometry, environment-based labeling, and
targeted on-site shifts provides a lightweight approach to modeling band
offsets and valley splittings without additional DFT calculations for each
superlattice period.  The monolayer-resolution sweep over 24 well widths
completes in under two minutes on a single core, illustrating the
efficiency of the approach for systematic parameter-space exploration.

\subsection{Twisted bilayer graphene}
\label{sec:tbg}

The most demanding application targets twisted bilayer graphene (TBG),
where the electronic structure depends sensitively on the interlayer
coupling, which varies continuously across the moir\'{e} unit cell as the
local stacking interpolates between AA and AB regions.  We construct an
EDTB model with $spd$ basis (9~orbitals/atom) fitted simultaneously to
nine training geometries: monolayer graphene, five AB-stacked bilayers at
interlayer separations $d = 3.00$, 3.20, 3.35, 3.50, and 3.70~\AA, and
three AA-stacked bilayers at $d = 3.00$, 3.35, and 3.70~\AA.  All DFT
calculations use PBE with a $18^2 \times 1$ $k$-grid.

\begin{figure}[t]
  \centering
  \includegraphics[width=0.85\columnwidth]{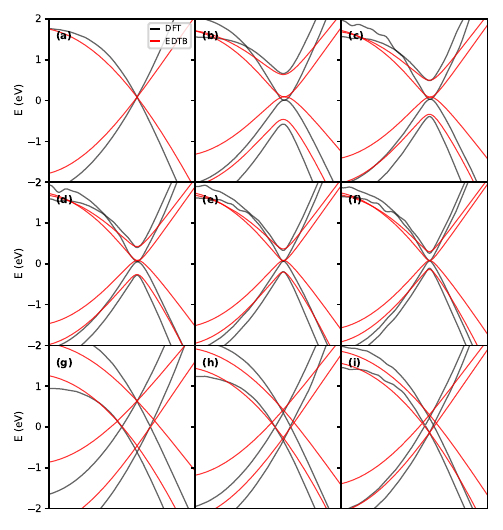}
  \caption{EDTB (red) vs DFT (black) near the $K$ point for the nine
    training geometries.
    (a)~Monolayer; (b--f)~AB bilayer; (g--i)~AA bilayer.}
  \label{fig:dirac_cone}
\end{figure}

The critical ingredient is the distance-dependent Goodwin hopping form
of Eq.~(\ref{eq:goodwin}) with reference distance $r_0 = 1.42$~\AA{}
and cutoff $r_c = 5.29$~\AA, which enables continuous interpolation
of the interlayer coupling.  The complete parameter set
(4~on-site energies, 10~amplitudes $V_{0,\lambda}$, 10~exponents
$n_\lambda$, 1~cutoff steepness $n_c$, 6~screening constants
$\gamma_{ll'}$---31 parameters) is determined by a simultaneous
Levenberg--Marquardt fit.  Besides some deviation away from the Dirac cone region, the model reproduces the
Dirac-cone dispersion, band splitting, and the distinct electronic
signatures of AB and AA stacking across the full range of interlayer
separations (Fig.~\ref{fig:dirac_cone}).

To demonstrate scalability, we compute band structures for three
commensurate TBG supercells: $(6,5)$ with $\theta = 6.01°$ (364~atoms),
$(10,9)$ with $\theta = 3.48°$ (1,084~atoms), and $(20,19)$ with
$\theta = 1.70°$ (4,324~atoms).  The corresponding Hilbert-space
dimensions are $\nawf = 3{,}276$, $9{,}756$, and $38{,}916$. In the three geometries the graphene atoms in the individual planes are not relaxed. 

Solving the eigenvalue problem for Hamiltonians of such large dimensions is not practical using traditional dense eigensolvers. We adopt a strategy that exploits the intrinsic sparsness of the EDTB Hamiltonian using on-the-fly
bond enumeration with the fitted distance-dependent and screening
parameters with no reparameterization required.
At each of the 80 $k$-points along the moir\'{e} 
Brillouin-zone path, the 40~eigenvalues nearest the shift-invert target
$\sigma = 0.0$~eV (Dirac point) are extracted via the implicitly restarted Lanczos
algorithm~\cite{Lanczos1950,Sorensen1992}.  This approach reduces the
memory footprint from $\sim$11~GB (dense) to $\sim$250~MB (sparse) for
the largest system.

\begin{figure*}[tp]
  \centering
  \includegraphics[width=\textwidth]{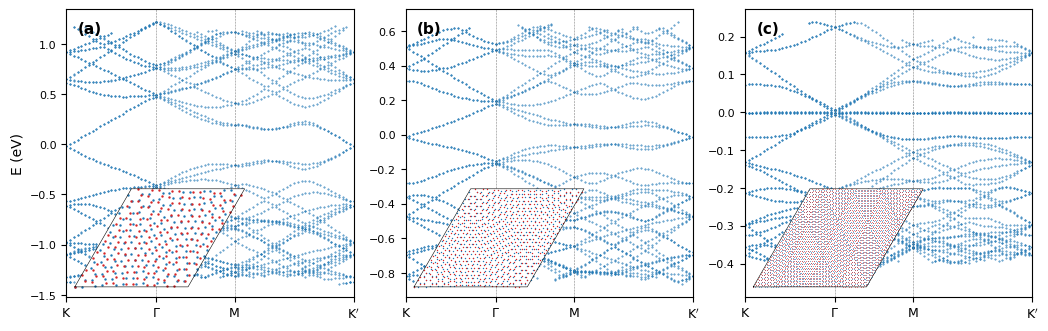}
  \caption{Sparse Lanczos band structure of TBG along
    $K$--$\Gamma$--$M$--$K'$ for three supercells:
    (a)~$(6,5)$, $\theta = 6.01°$, 364~atoms;
    (b)~$(10,9)$, $\theta = 3.48°$, 1,084~atoms;
    (c)~$(20,19)$, $\theta = 1.70°$, 4,324~atoms.
    40~eigenvalues nearest $\sigma = 0.0$~eV are plotted per
    $k$-point.  Insets show the moir\'{e} unit cell.}
  \label{fig:tbg_bands}
\end{figure*}

Figure~\ref{fig:tbg_bands} shows the results.  For the $(6,5)$
supercell, folded Dirac cones are clearly resolved.  At
$\theta = 3.48°$ [$(10,9)$], increased interlayer hybridization
produces a dense manifold of flat bands with reduced bandwidth.  For the
$(20,19)$ supercell at $\theta = 1.70°$---approaching the magic
angle---the spectrum near the Fermi level is dominated by nearly flat
bands with bandwidth $\sim$10--20~meV, consistent with the expected
flattening of the low-energy spectrum in the small-angle
regime~\cite{BistrMacD2011}.

The scalability of the sparse EDTB approach is summarized in
Table~\ref{tab:tbg_scaling}.  The sparsity of the Hamiltonian---arising
from the finite hopping cutoff---yields memory and time scaling that is
approximately linear in $\Nat$ for fixed numbers of target eigenvalues,
making systems with $\nawf \sim 10^5$ accessible on a single workstation.

\begin{table}[t]
\centering
\caption{TBG supercell sizes and memory requirements.}
\label{tab:tbg_scaling}
\begin{tabular}{crrrrr}
  \toprule
  $(n,m)$ & $\theta$ (deg) & $\Nat$ & $\nawf$ &
    Sparse (MB) & Dense (GB) \\
  \midrule
  $(3,2)$   & 13.17 &     76 &     684 &   0.1 &  0.004 \\
  $(6,5)$   &  6.01 &    364 &  3,276 &   2   &  0.08  \\
  $(10,9)$  &  3.48 &  1,084 &  9,756 &   6   &  0.7   \\
  $(20,19)$ &  1.70 &  4,324 & 38,916 &  25   & 11     \\
  \bottomrule
\end{tabular}
\end{table}


\section{Conclusions}
\label{sec:conclusions}

We have presented a comprehensive framework for constructing
environment-dependent tight-binding models from \textit{ab initio}
pseudo-atomic orbital Hamiltonians.  The method builds upon the
Slater--Koster two-center approximation, augmented with bond-screening
functions that capture the influence of the local atomic environment on
the electronic structure.  By fitting to the eigenvalue spectrum of the
PAO Hamiltonian, our approach avoids the gauge ambiguities
inherent in Wannier-function-based downfolding while preserving the
symmetry and orbital character of the \textit{ab initio} representation.
This method enables large-scale simulations of the electronic, optical and transport properties of materials systems with DFT accuracy at a fraction of the computational cost.

\section{Data availability}
  The complete EDTB fitting and Hamiltonian construction code is
implemented as a module within the \paoflow{} package at \url{https://github.com/marcobn/PAOFLOW} (and \texttt{pip install paoflow}) and the code needed to reproduce the presented results is available
at \url{https://github.com/marcobn/EDTB_code}.

\vspace{1em}


\begin{acknowledgments}

MBN wishes to thank Marco Fornari, Marcio Costa, Jagoda Sławińska, Anooja Jayaraj and Davide Ceresoli for enlightening and stimulating discussions. This work has been supported, in part, by the US Department of Energy (DOE, Office of Basic Energy Sciences) under grant No. DE-SC0024554, and the National Science Foundation under grant No. DMR2523217.
Portions of the code development, debugging, and documentation for this
work were carried out with the assistance of GitHub Copilot (Claude Opus
4.6~\cite{Anthropic2025Claude4}, Anthropic), an AI-powered programming assistant integrated into the
Visual Studio Code development environment. All scientific decisions---including model design,
choice of physical approximations, parameter selection, 
interpretation of results, and writing of the paper---were made by the author.
\end{acknowledgments}

\FloatBarrier

\bibliography{PAOFLOW_EDTB_paper}

\end{document}